\pdfoutput=1

\documentclass[12pt]{iopart}

\usepackage{graphicx}
\usepackage{siunitx}
\usepackage{xcolor}
\usepackage{subcaption}
\usepackage{adjustbox}
\usepackage{setspace}

\usepackage{hyperref}

\begin{document}
\title[]{Hands-on Experiment Supported by Augmented Reality Smartglasses for Learning the Lorentz Force}
\author{Max Warkentin$^1$, Kristin Altmeyer$^2$, Yajie Liang$^1$, Bermann Steinmacher$^3$, Barbara Gränz$^3$, Andreas Lichtenberger$^3$, Stefan Küchemann$^1$, Jochen Kuhn$^1$ and Christoph Hoyer$^1$}
\address{$^1$ Faculty of Physics, Ludwig-Maximilians-Universität München (LMU Munich), 80539 Munich, Germany} 
\address{$^2$ Department of Education, Saarland University, 66123 Saarbrücken, Germany}
\address{$^3$ Laboratory for Solid State Physics, ETH Zurich, 8093 Zurich, Switzerland}

\ead{max.warkentin@physik.lmu.de}

\begin{abstract}
Previous research has shown that inquiry-based learning through hands-on experiments, as well as learning with multiple, external representations (MERs) can promote the understanding of complex phenomena in physics. In this context, augmented reality (AR) smartglasses make it possible to superimpose real experiments with virtually presented information while maintaining learners’ freedom to interact with the experimental setup. This allows spatial and temporal contiguity to be established between the phenomenon and the supporting virtual visualizations which can help to reduce learners' cognitive load and to free up cognitive resources for learning. We introduce an AR learning environment for the investigation of the Lorentz force. By visualizing the fields relevant to the experiment in AR, a more direct investigation of the physical phenomenon can be achieved for the learner. Additionally, this setup can foster the acquisition of representational skills by providing the possibility to superimpose fields using various representations. This learning setting will be used in future studies to investigate how MERs should be combined in a guided inquiry-based learning environment to best promote learners' conceptual knowledge and representational skills. We present the system and report usability data that were collected from a sample of 188 secondary school students. The average score was 80.18 (SD: 12.06) which signals an excellent usability of the AR learning environment.
\end{abstract}

\section{Introduction}
In physics education, active learning and hands-on experiments have long been known to be beneficial for learners \cite{chi2014icap}. However, more recent research has pointed out that the benefits of laboratory courses for the acquisition of conceptual knowledge are not guaranteed \cite{PhysRevX.10.011029} and has demonstrated that the specific implementation of such an inquiry-based learning environment is critical \cite{PhysRevPhysEducRes.13.010129,worner2022best}. 

One of the challenges that physics learners often face is the fact that the phenomena they are investigating while experimenting are not accessible to direct observation. This is especially true when dealing with electromagnetism. There, the electric and magnetic fields are important agents that are responsible for a multitude of phenomena. However, students cannot see these entities.

Hence, a powerful way to support learners would be to provide real-time, spatially aligned visual representations of the electric and magnetic fields while they conduct hands-on experiments. This can be achieved with augmented reality (AR) devices, such as AR smartglasses \cite{10.1162/pres.1997.6.4.355}, which overlay the user’s field of view with digital content---thus, for example, making an otherwise invisible magnetic field visible. It has already been shown in a number of previous studies that AR can foster conceptual understanding in physics \cite{garzon2019meta,altmeyer2020use,malone2023three}. A potent feature of AR learning environments is that supporting visualizations can be displayed spatially and temporally close to the physical phenomenon, thereby reducing the learner’s cognitive load \cite{sweller1998cognitive,thees2020effects}. 

However, previous research has also revealed that not all AR learning environments increase learning gains and that the specific implementation of AR matters \cite{de2019moving,husnaini2019effects}. Furthermore, although many studies have investigated the learning effects of general AR devices, such as tablets and smartphones, the learning effectiveness of the AR smartglasses technology, in particular, is still not well researched \cite{laumann2024analyzing}. Meanwhile, these AR devices bear great potential for hands-on experiments, because they keep the users' hands free to interact with the experimental components while possible changes in the digitally provided visualizations or measurement values can be observed simultaneously. Thus, the interaction with natural phenomena is even more direct than in non-AR experiments \cite{altmeyer2020use}. Additionally, learners can be supported in the measurement process by either fully automated or highly assisted data acquisition, so that sufficient cognitive resources are still available for examining the representations of the measured values \cite{sweller1998cognitive}.

In order to guide the visualizations of various phenomena in physics, a plethora of representations has been invented. For example, in the case of electromagnetism, there are multiple ways to visualize the magnetic field, each of them stressing a slightly different facet. It is known that such external representations can help to connect real phenomena to physics concepts \cite{olympiou2013making} and that it can be even more beneficial to use multiple external representations (MERs) (see e.g.\ \cite{mayer1997multimedia}) to combine the advantages of different representations \cite{ainsworth2006deft}. However, the proper understanding and application of representations requires a certain minimum of representational competence and the acquisition of conceptional knowledge could be influenced by this competence \cite{rau2017conditions,hubber2010teaching,nitz2014student}. An increased number of representations bears the potential of increasing the learners' cognitive load \cite{schroeder2018spatial} but it can foster learning, too \cite{rexigel2024more}. Thus, taking these considerations into account, it is of utmost importance to support the learners in constructing representational competence.

In the present work, we introduce a learning environment for studying the Lorentz force in a setting that involves an AR-enhanced hands-on experiment with carefully chosen MERs. The AR learning environment is suitable for both high school students and university level (physics) students. It can be used in a laboratory course setting to facilitate the students' understanding of how the Lorentz force on a current carrying conductor relates to various manipulations of the current and the orientation of the external magnetic field, by showing representations of the magnetic fields and other relevant quantities in three dimensions.\footnote{For an early proposal of this AR setup, see \cite{donhauser2020making}.} This opens up promising possibilities for an easier access to the topic of the Lorentz force, as well as the understanding of complex magnetic field structures.

This paper is organized in the following way. In \sref{sec:learning_content}, we review the learning content that we wish to address with the learning environment and explain the choice of our visualizations. In \sref{sec:calculations}, we describe how the representations are produced. In \sref{sec:ar_setup}, we introduce the AR learning environment and describe its 3D visualizations and functionality. We also report the data from the evaluation of the usability of the system. In \sref{sec:conclusion}, we summarize the work and outline possible future extensions of this AR setup.

\section{Learning Content}  \label{sec:learning_content}
\subsection{The Conductor Swing Experiment} \label{sec:classical_experiment}
One of the most important fundamental interactions in physics is the electromagnetic interaction between photons and electrically charged particles. The force that is experienced by a charged particle in the presence of electric or magnetic fields, is called the \emph{Lorentz force}. When there is no electric field present, the particle is only subject to the magnetic force.\footnote{Frequently, the term ``Lorentz force'' is used for this force alone. We will adhere to this custom.}

In order to investigate this force experimentally, it is often more convenient to expose a conductor, carrying a steady electric current with magnitude $I$, to the magnetic field $\mathbf{B}$. In the idealized case of an infinitesimally thin conductor, the Lorentz force on an infinitesimal segment of that conductor is given by
\begin{equation} \label{eq:lorentz_force_mag_current}
\rmd \mathbf{F}_{\rm{mag}} = I \rmd \mathbf{l} \times \mathbf{B}\,,
\end{equation}
where $\rmd \mathbf{l}$ denotes the orientation and (infinitesimal) length of that segment.

A popular experimental setup that probes relation \eref{eq:lorentz_force_mag_current} and supports students in learning how the Lorentz force, and particularly its direction, depends on the relevant quantities, is the \emph{conductor swing experiment}. 
\begin{figure}[h]
 \centering
  \includegraphics[width=12cm]
      {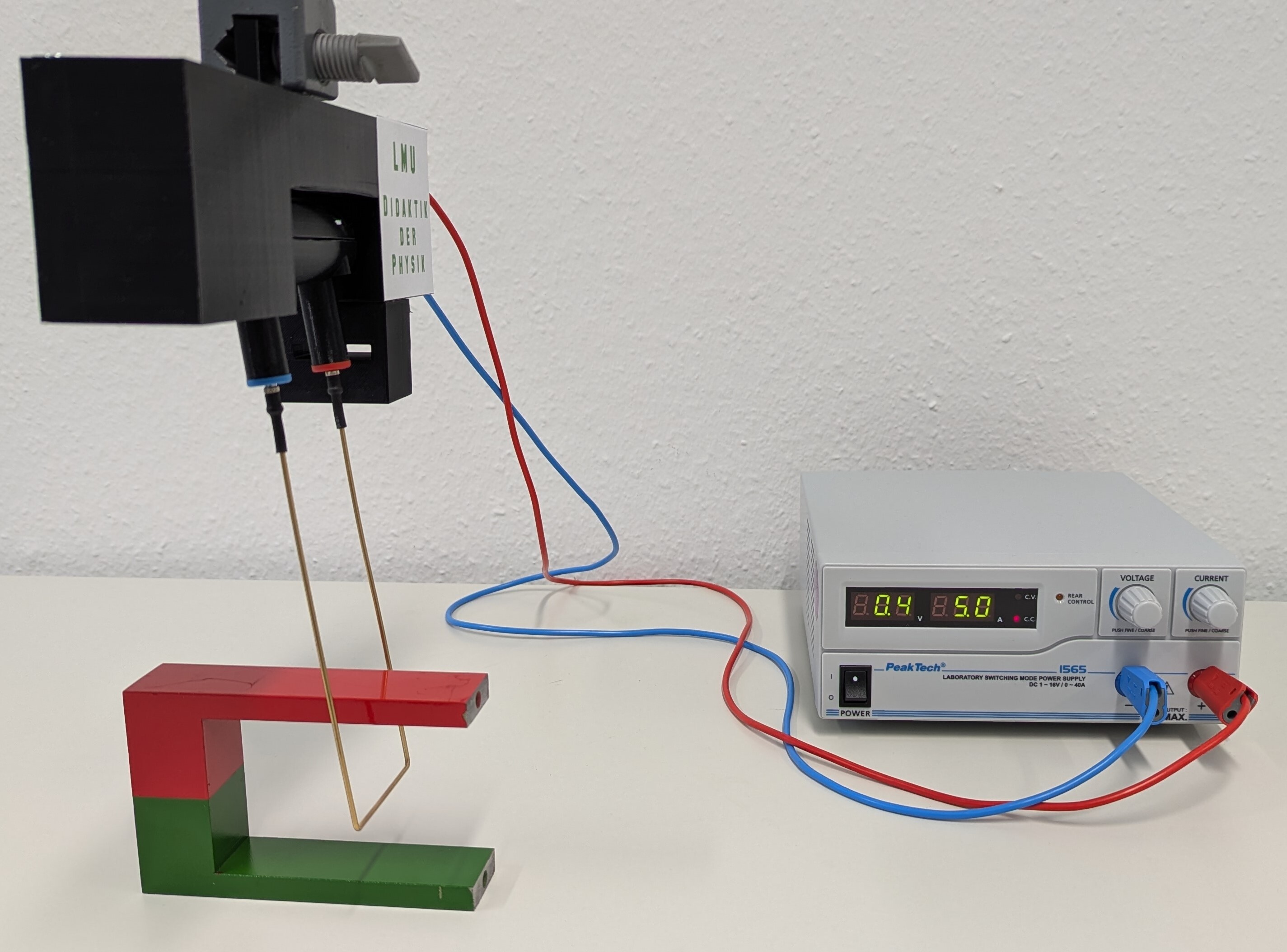}
   \caption
   {Typical setup of the conductor swing experiment, with the horseshoe magnet, conductor, conductor mounting and power supply.}
   \label{fig:trad_setup}
\end{figure}
As can be seen in \fref{fig:trad_setup}, in this setup a swing-shaped conductor passes through the interior of a horseshoe-shaped permanent magnet that provides the external magnetic field. The Lorentz force due to the directions of the current and the magnetic field of the horseshoe magnet causes the conductor to swing into one of two directions. The learners can interact with that physical system by adjusting the current through the conductor and by changing the direction of the external magnetic field, by flipping or rotating the magnet. They can then observe how the direction of the Lorentz force depends on these variations.

\subsection{The Magnetic Fields} \label{sec:mag_fields}
The external magnetic field is provided by a horseshoe-shaped permanent magnet (see \fref{fig:trad_setup}), which consists of a central piece that is a bar magnet and side flanks that are made of soft magnetic materials. Because of the complex interaction of the permanent magnet with the magnetizable soft magnetic materials, the magnetic field of the full horseshoe magnet cannot be calculated analytically. For the numerical calculation, the Python library \textsc{Magpylib} \cite{ORTNER2020100466} and its extension \textsc{Magpylib-Material-Response}\footnote{\url{https://magpylib-material-response.readthedocs.io/en/latest}} can be used. The first package provides powerful numerical methods to derive the magnetic field based on analytical solutions for the magnetic fields of rectangular-shaped permanent magnets. The second package models the material response by using the magnetostatic method of moments \cite{1608257}. 

Another magnetic field that is present in the system is the field that is generated by the electric current through the conductor. Considering the ideal case of an infinitely thin conductor, its magnetic field can be simply calculated by applying the Biot-Savart law or, conveniently, by using the aforementioned Python packages.

We use these numerical methods to determine the magnetic fields of the horseshoe magnet and the conductor in \sref{sec:exact_external_field} and \sref{sec:exact_conductor_field}, respectively.

\subsection{Superposition of Magnetic Fields} \label{sec:superposition}
An important concept in electromagnetism is the (linear) superposition of magnetic fields. For example, in order to determine the magnetic force that acts on a moving, electrically charged particle, one has to add all magnetic fields that are present at the particle's position. Aside from the conceptual importance, if students learn to superimpose and split up magnetic fields, it can also foster their representational competence of vector fields \cite{kuchemann2021inventory}.

In the Lorentz force experiment, we can consider the superposition of the external magnetic field of the horseshoe magnet and the magnetic field produced by the conductor if it carries an electric current. Besides the possible gain for representational competence, another advantage is that visualizing this superposition provides additional information that can be used to deduce the direction of the current and thus the direction of the Lorentz force. This may offer an alternative approach for learners that struggle with the right-hand rule representation (see \sref{sec:rep_tripod}) that relates to the mathematical concept of the vector product and hence might be difficult to understand for less knowledgeable learners, especially at the level of high school students.

\subsection{Representations} \label{sec:reps}
In the present section, we will describe the three representations that are provided in our learning environment:
\begin{enumerate}
    \item Field vector representation
    \item Field line representation
    \item Right-hand rule representation
\end{enumerate}

Two of these representations are supposed to make the invisible magnetic fields visible---the field vector and field line representations, respectively. These representations are very common in physics education, although they are usually applied for 2D visualizations on a plane. The enormous advantage of our AR learning environment is that we can utilize these representations directly in three dimensions where they most closely model the actual 3D distribution of the magnetic fields. In the case of the field line representation, all three dimensions are even necessary to turn this representation from a mere qualitative one into a quantitatively correct picture \cite{wolf1996electric}.\footnote{An extensive overview of misconceptions regarding field vector representations and their connection to field line representations is also provided in \cite{hoyer2024vector}.}

The third representation that we include---the right-hand rule representation---is useful in visualizing the relevant physical quantities for the context of the Lorentz force. 

From a didactic-pedagogical perspective, these three representations can be understood as scientific model representations used to understand, explain, or predict phenomena and processes \cite{passmore2013models}. According to Ainsworth's DeFT framework \cite{ainsworth2006deft}, the combination of multiple (model) representations in learning contexts can serve different functions: they can complement each other, constrain each other’s interpretation, or promote the construction of deeper understanding. In the current physics lab-work environment, two of these functions are addressed. First, the AR-based integration of model representations complements the real-world representation of the experimental setup by visualizing underlying principles and mechanisms. Second, the current learning environment encourages learners to mentally connect the three model representations and can thereby promote a deeper understanding of electromagnetic concepts.

\subsubsection{The Field Vector Representation} \label{sec:rep_fieldvector}
The field vector representation emphasizes the vector field character of the magnetic field. At arbitrarily chosen locations, a representative vector is drawn that encodes the value of the magnetic field at that point by the use of its length and orientation. For this, we need to choose a grid of visualization points, which is usually chosen with equidistant points, and calculate the magnetic field at those points. The correspondence between a particular field strength and a respective arrow length is arbitrary and thus can be chosen freely. However, after that initial choice all other arrow lengths have to scale according to the relation of the field strengths.

\subsubsection{The Field Line Representation}
The field line representation makes it particularly simple to visualize how a collection of tiny magnetic dipoles would be oriented in a magnetic field by displaying the field lines that would be tangent to the directions of the dipoles. Additionally, the direction of the magnetic field along any field line is provided by displaying arrows along the lines. Another feature of this representation is that the relative strengths of certain regions of the magnetic field can be encoded by the relative spacings of the field lines, i.e.\ the field line density. Regions with denser magnetic field lines then represent larger magnetic field strengths while regions with a lower field line density correspond to weaker magnetic fields. 

Unfortunately, this particular feature makes it typically more complicated to produce a proper field line representation. Since the density of field lines should be an indicator of the field strength, one cannot usually choose equidistant points as seed points for the field lines but has to choose a distribution of seed points that mirrors the behavior of the field strength. We explain how this can be achieved in \sref{sec:calc_fieldlines}.

\subsubsection{The Right-Hand Rule Representation} \label{sec:rep_tripod}
\begin{figure}[h]
 \centering
  \includegraphics[width=8cm]
      {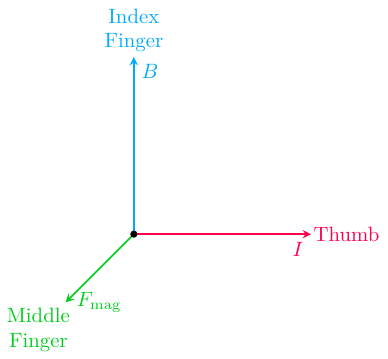}
   \caption
   {Right-hand rule representation. The axes are labeled with the physical quantities and with the associated fingers of the right hand.}
   \label{fig:tripod_2d}
\end{figure}

The right-hand rule representation is not intended to support visualizing the magnetic field itself but rather visualizes that the Lorentz force $\mathbf{F}_{\rm{mag}}$ is always orthogonal to the electric current $\mathbf{I}$ and the magnetic field $\mathbf{B}$, in addition to providing the correct direction. The underlying physics is encoded in \eref{eq:lorentz_force_mag_current} by means of the vector product. The latter is represented by showing orthogonal arrows, distributed in a right-handed configuration, in three different colors corresponding to the quantities $\mathbf{I}$, $\mathbf{B}$ and $\mathbf{F}_{\rm{mag}}$. Additionally, the representation encourages to use ones right hand to arrange those vectors in the correct way, by associating them with the thumb, index finger and middle finger. \Fref{fig:tripod_2d} displays this representation.

\section{Visualization of the Magnetic Fields} \label{sec:calculations}
\subsection{Modeling the Magnetic Field of the Rectangular Horseshoe Magnet} \label{sec:exact_external_field}
In the following, we will determine the magnetic field of a horseshoe magnet like the one visible in \fref{fig:trad_setup}. According to the data sheet, the relative permeability and the remanence of the (central) AlNiCo permanent magnet are 
\begin{equation*}
\mu_r = 3 \quad \textrm{and} \quad J = \qty{1.25}{T} \qquad (\textrm{magnet})\,,
\end{equation*}
respectively. This magnet magnetizes the attached soft magnetic pieces. Although we have no knowledge about the exact materials that are used for the latter, typical values of $\mu_r$ for soft magnetic iron are in the range of $\sim$4000 to $\sim$6000. Since one can verify that the particular value does not affect the resulting magnetic field significantly, in the following, we will use the value
\begin{equation*}
\mu_r = 5000 \qquad (\textrm{iron})\,.
\end{equation*}

\begin{figure}[h]
 \centering
  \includegraphics[width=15cm]
      {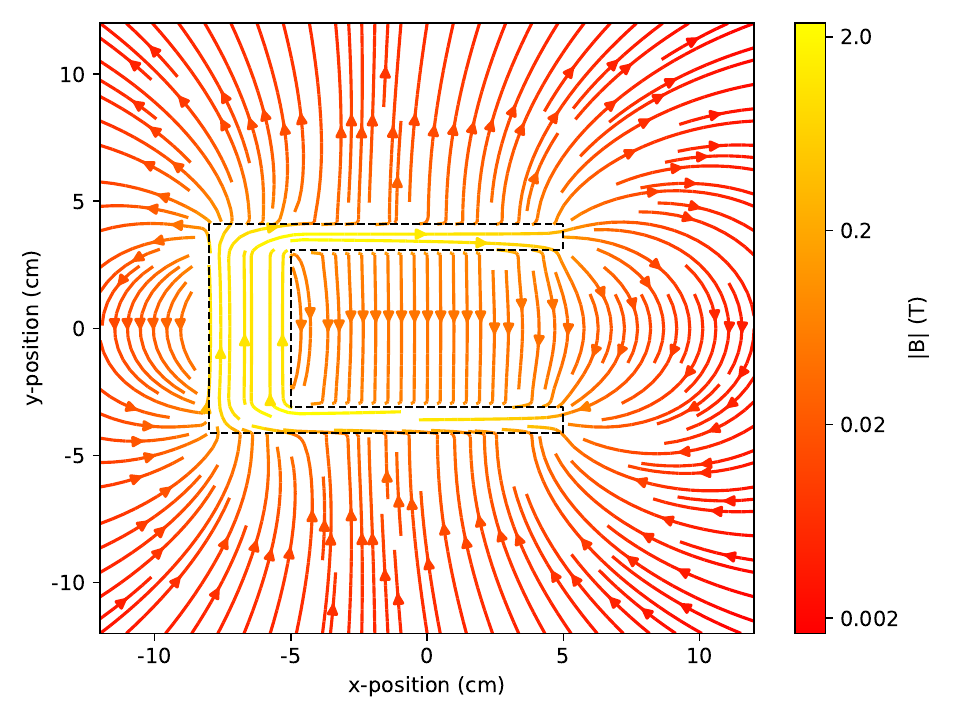}
   \caption
   {2D (cross-sectional) streamline plot of the magnetic field of the horseshoe magnet. The strength of the magnetic field is (logarithmically) encoded by color. The shape of the horseshoe magnet is superimposed on the plot by using the dashed outline.}
   \label{fig:streamplot_horseshoe_2d}
\end{figure}

\begin{figure}[h]
 \centering
  \includegraphics[width=15cm]
      {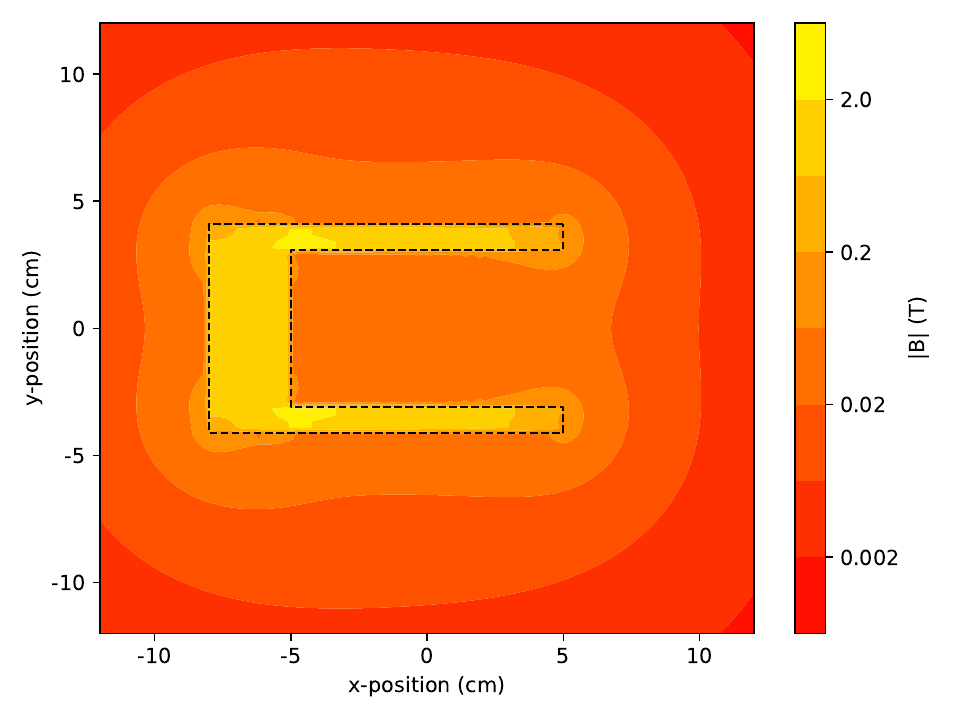}
   \caption
   {2D (cross-sectional) filled contour plot of the magnetic field of the horseshoe magnet. The strength of the magnetic field is (logarithmically) encoded by color. The shape of the horseshoe magnet is superimposed on the plot by using the dashed outline.}
   \label{fig:contourplot_horseshoe_2d}
\end{figure}

Using the Python libraries introduced in \sref{sec:mag_fields}, we can calculate the magnetic field of the horseshoe magnet. The result is visualized in figures \ref{fig:streamplot_horseshoe_2d} and \ref{fig:contourplot_horseshoe_2d}.

It is apparent from these figures that it is reasonable to consider the interior of the horseshoe magnet to be approximately homogeneous, as long as one stays away from the edges. 

Using the value for the magnetic field strength in the center (where the conductor is at rest), we roughly estimate
\begin{equation*}
B_{0,\, \rm{th}} \simeq \qty{46}{mT}\,.
\end{equation*}

The real value, which we obtained through a magnetic field measurement, is slightly lower,\footnote{There are several factors that can contribute to this discrepancy, including a steady demagnetization of permanent magnets over time or that the actual material specification slightly deviates from the parameters that we used.}
\begin{equation} \label{eq:b_0_exp}
B_{0,\, \rm{exp}} \simeq \qty[separate-uncertainty-units = bracket]{32.10(5)}{mT}\,.
\end{equation} 

Thus, in the present work, for producing the magnetic field visualizations between the two legs of the horseshoe magnet we use a homogeneous magnetic field, given by
\begin{equation} \label{eq:mag_field_ext}
\mathbf{B}_{\rm{ext}} = B_{0,\, \rm{exp}} \mathbf{n}\,,
\end{equation}
where the unit vector $\mathbf{n}$ points in the direction from the upper flank (red-colored side of the magnet in \fref{fig:trad_setup}) to the lower flank (green-colored side) and $B_{0,\, \rm{exp}}$ is given by \eref{eq:b_0_exp}.

\subsection{Modeling the Magnetic Field of the Conductor Swing} \label{sec:exact_conductor_field}
\begin{figure}[h]
 \centering
  \includegraphics[width=\textwidth]
      {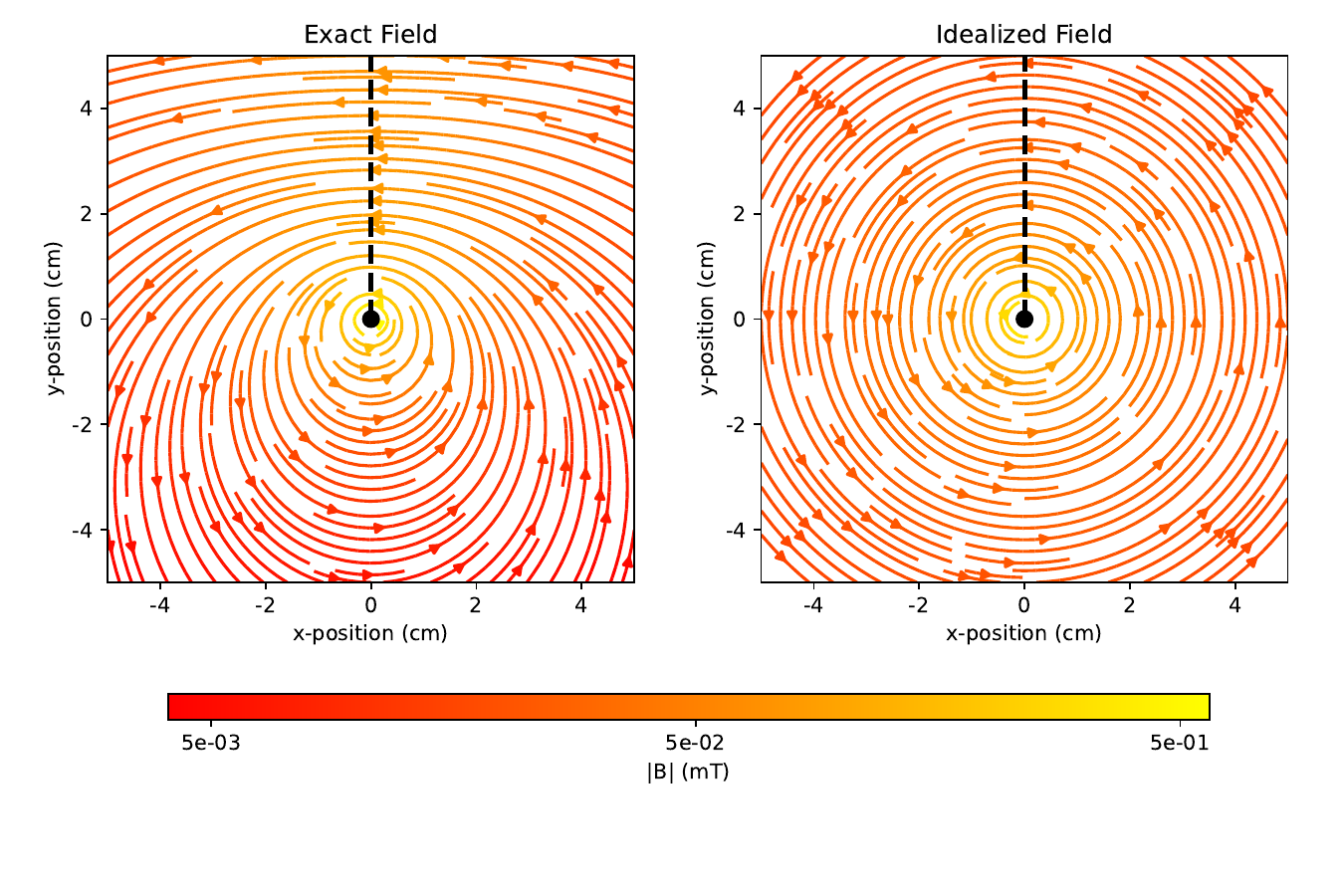}
   \caption
   {2D streamline plots of the magnetic field of the conductor, in a plane orthogonal to the horizontal conductor and centered in the middle. $I = \qty{5}{A}$, which is a typical value that is used during the experiment. The strength of the magnetic field is (logarithmically) encoded by color. The shape of the conductor, as viewed from the side, is indicated in the plots by using the black outlines. The left panel shows the magnetic field due to all three conductor swing segments. The right panel shows the field due to an infinitely long horizontal conductor.}
   \label{fig:streamplots_conductor_2d}
\end{figure}

The conductor swing consists of a horizontal segment with length $\qty{6}{cm}$ and two vertical segments that have a length of $\qty{16}{cm}$, each. The magnetic field depends on the direction and magnitude of the electric current. The left panel of \fref{fig:streamplots_conductor_2d} 
shows the exact magnetic field, when all three segments contribute to the field. We used the Python package \textsc{Magpylib} for the calculation (see \sref{sec:mag_fields}). The right panel displays an idealized magnetic field due to an infinitely long horizontal conductor, where the vertical segments become irrelevant at the location of the horseshoe magnet. As can be seen in \fref{fig:streamplots_conductor_2d}, the exact magnetic field is distorted with respect to the idealized one, because the currents going through the vertical segments contribute to the magnetic field. 

However, since this circumstance is due to the geometry of our conductor, it can be diminished by making the horizontal segment longer. Thus, for the sake of simplicity, in our AR setup we consider the conductor to be an infinitely long horizontally oriented wire. In this case, the magnetic field is given by (using the Biot-Savart law)
\begin{equation} \label{eq:mag_field_cond}
\mathbf{B}_{\rm{cond}} = \frac{\mu_0 I}{2 \pi s^2} \left(- y, x, 0\right)^{\rm{T}}\,, \qquad s = \sqrt{x^2 + y^2} \,,  
\end{equation}
with $x$ and $y$ spanning the plane orthogonal to the conductor and the $z$-axis pointing along the conductor (positive current direction points towards the viewer).

\subsection{Visualizing the Superposition of Magnetic Fields} \label{sec:visualization_superposition}
When we compare the strength of the conductor's magnetic field
\begin{equation*} 
|\mathbf{B}_{\rm{cond}}(y=0)| = \qty{0.1}{mT} \frac{I/(\qty{5}{A})}{x/\unit{cm}}\,,
\end{equation*}
evaluated along the distance $x$ from the conductor, to the strength of the external magnetic field 
\begin{equation*} 
|\mathbf{B}_{\rm{ext}}| = \qty{32}{mT}\,,
\end{equation*}
where we used $B_{0,\, \rm{exp}}$ from \eref{eq:b_0_exp}, the ratio can be expressed as
\begin{eqnarray*}
\frac{|\mathbf{B}_{\rm{cond}}(y=0)|}{|\mathbf{B}_{\rm{ext}}|} \simeq 0.003 \frac{I/(\qty{5}{A})}{x/\unit{cm}}\,.
\end{eqnarray*}
Thus, it can be seen that the two magnetic field strengths are only comparable for either very large currents or very small distances from the conductor. 

However, we decided to include the option to artificially exaggerate the effect of the conductor (i.e.\,, effectively turning up the current) in the AR learning environment, so that both magnetic fields could have a comparable strength in the vicinity of the conductor. In this way, learners can interact more extensively with the visualizations provided by the AR learning environment, thereby increasing their opportunity to learn about the Lorentz force and improve their representational competence, as we explained in \sref{sec:superposition}.

Thus, the third magnetic field that is visualized in the AR setup is the total magnetic field
\begin{equation} \label{eq:mag_field_total}
\mathbf{B}_{\rm{total}} = \mathbf{B}_{\rm{ext}} + \mathbf{B}_{\rm{cond}} 
\end{equation}
due to the superposition of the external and the conductor's magnetic fields.

\subsection{Calculation of Magnetic Field lines} \label{sec:calc_fieldlines}
The first step for generating the field lines is to find suitable seed points. To simplify this task, it is sometimes possible to choose the seed points in a region where the magnetic field is homogeneous so that these points can be chosen equidistantly. This is trivial for the case of the homogeneous external field $\mathbf{B}_{\rm{ext}}$. 

\begin{figure}[h]
 \centering
  \includegraphics[width=12cm]
      {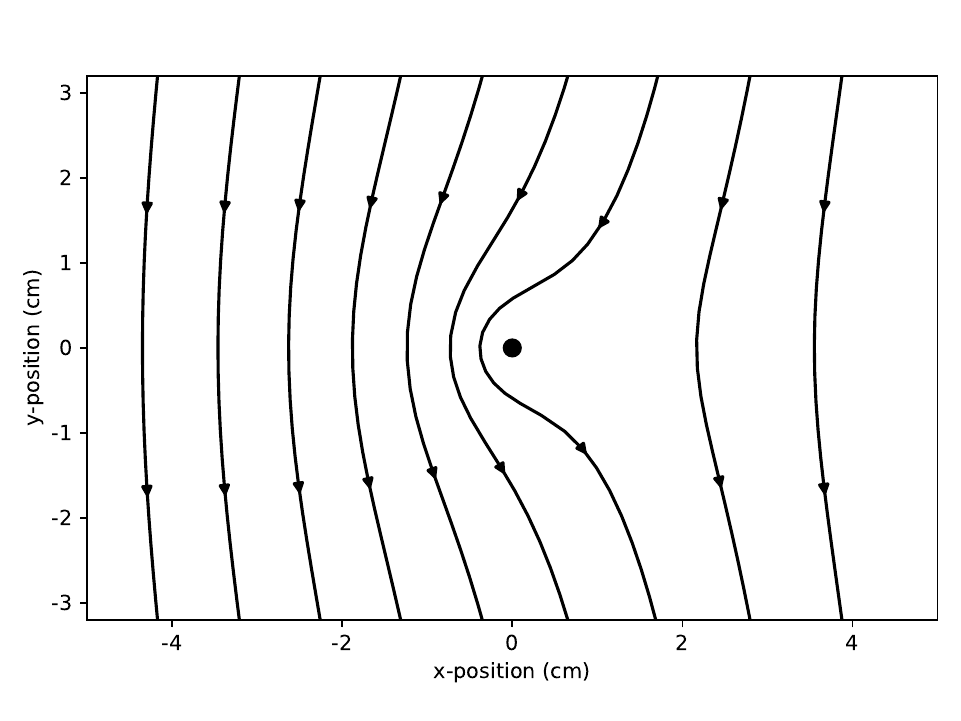}
   \caption
   {2D field line plot of the total magnetic field $\mathbf{B}_{\rm{total}}$ using the amplified value $I = \qty{1500}{A}$. The plotting range corresponds to the region of the interior of the horseshoe magnet. The shape of the conductor, as viewed from the side, is indicated as a thick dot.}
   \label{fig:fieldlineplot_superposition_2d}
\end{figure}

Also, for the total magnetic field $\mathbf{B}_{\rm{total}}$, we can make use of the fact that away from the conductor the total magnetic field approaches the homogeneous magnetic field, so that we can again use equidistant seed points. In \fref{fig:fieldlineplot_superposition_2d}, this method was used (there, the seed points are located a distance of $\qty{1}{m}$ away from the conductor).
Although in principle, there could also be some field lines that are closed around the conductor and hence are not seeded far away from it, these are present only for regions very close to the conductor. In our setup, we do not visualize such field lines to keep the visualizations comprehensible.

For the conductor's magnetic field we can use the following approach to find the positions of the seed points on the $x$-axis (at $y=0$). We equate the density of these points to the magnetic field \eref{eq:mag_field_cond} and choose some desired number of total points as the normalization constant. Then, we integrate along the $x$-axis (for $x>x_0$, where $x_0$ is some cutoff distance from the conductor) to determine the intervals that contain exactly one seed point. We can then, for simplicity, put one seed point in the center of each interval.\footnote{This method becomes more and more exact for larger and larger numbers of intervals.} 

Finally, we use the second-order Runge-Kutta method to trace out the field lines starting from those seed points. With this iterative method, it is possible to approximately follow the exact field line by estimating the direction of the magnetic field at consecutive points along the field line.

\section{The AR Learning Environment} \label{sec:ar_setup}
\subsection{Description of the AR Application}
We extended the experimental setup described in \sref{sec:classical_experiment} with the help of AR. In particular, we developed an application, written with the Unity software,\footnote{\url{https://unity.com}} that runs on the Microsoft HoloLens 2 (HL2).\footnote{\url{https://learn.microsoft.com/en-us/hololens/hololens2-hardware}} It digitally displays the 3D versions of the representations that were already introduced in \sref{sec:reps} in the physical experimental setup. In this way, the learners perceive information that is not present in the real world. The field vector and field line representations can be selected to visualize either the magnetic field of the permanent magnet, the magnetic field of the conductor, or their superposition. If the representation of the right-hand rule is selected, it does not vary with the selected field but is switched off for vanishing current.

\setcounter{footnote}{0}
\subsubsection{Adaptive Visualizations in Real Time}
\begin{figure}[h]
 \centering
  \includegraphics[width=10cm]
      {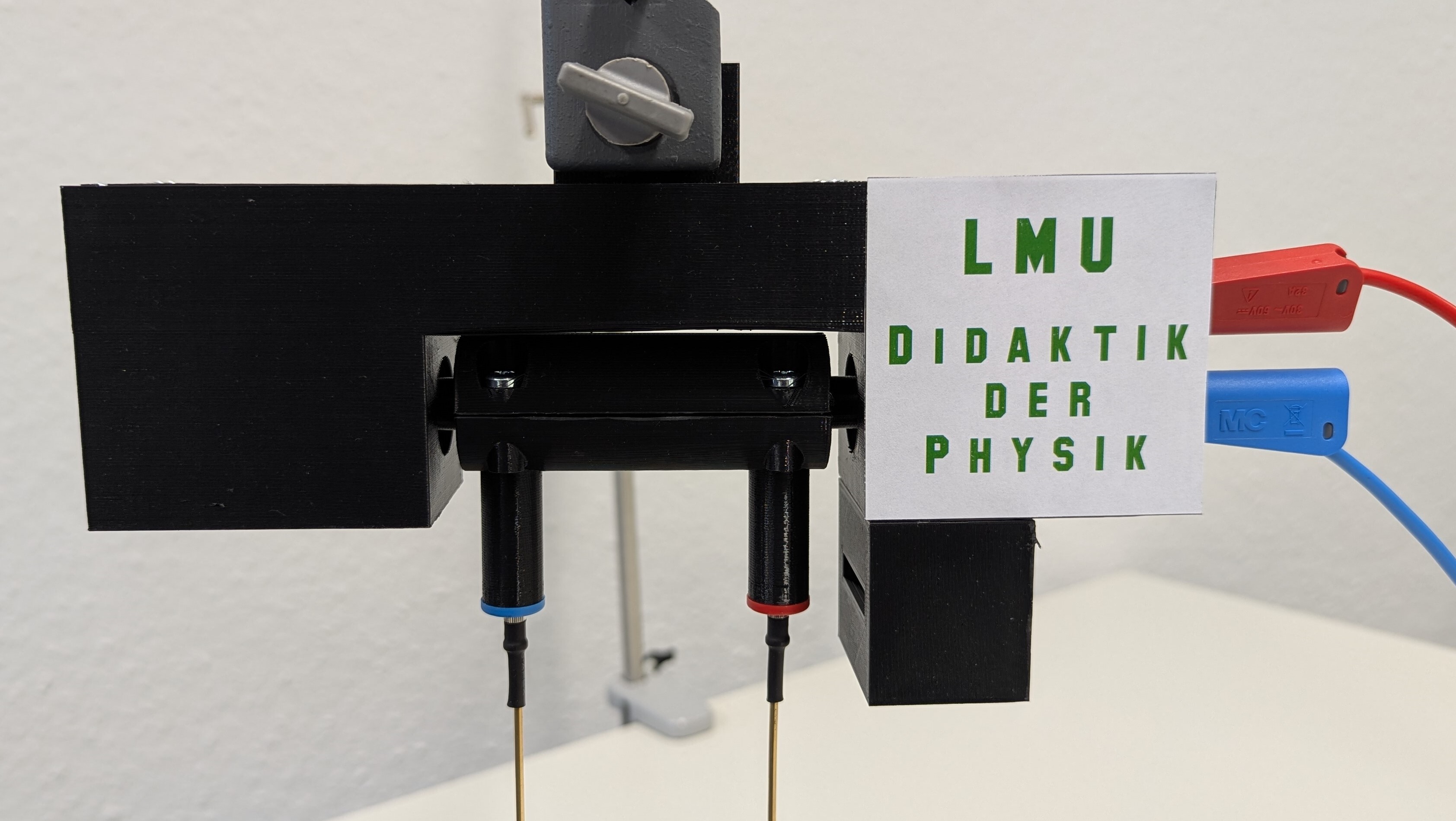}
   \caption
   {The conductor mounting containing the necessary electronics such as an electric current sensor, a rotary encoder and an Arduino microcontroller.}
   \label{fig:conductorMounting}
\end{figure}

For the representations to be effective in supporting the learners, it is crucial that they are displayed at exactly the right locations in the experimental setup and that they change according to the learners' manipulations. For the real-time measurement of the current and the conductor displacement, we have built an electric circuit including a current measurement sensor, a rotary encoder and an Arduino microcontroller with Bluetooth capability. All these components are integrated in the conductor mounting (see \fref{fig:conductorMounting}).\footnote{The specifics of this construction and the involved processes will be reported in detail, separately.}

\begin{figure}[h]
 \centering
  \includegraphics[width=10cm]
      {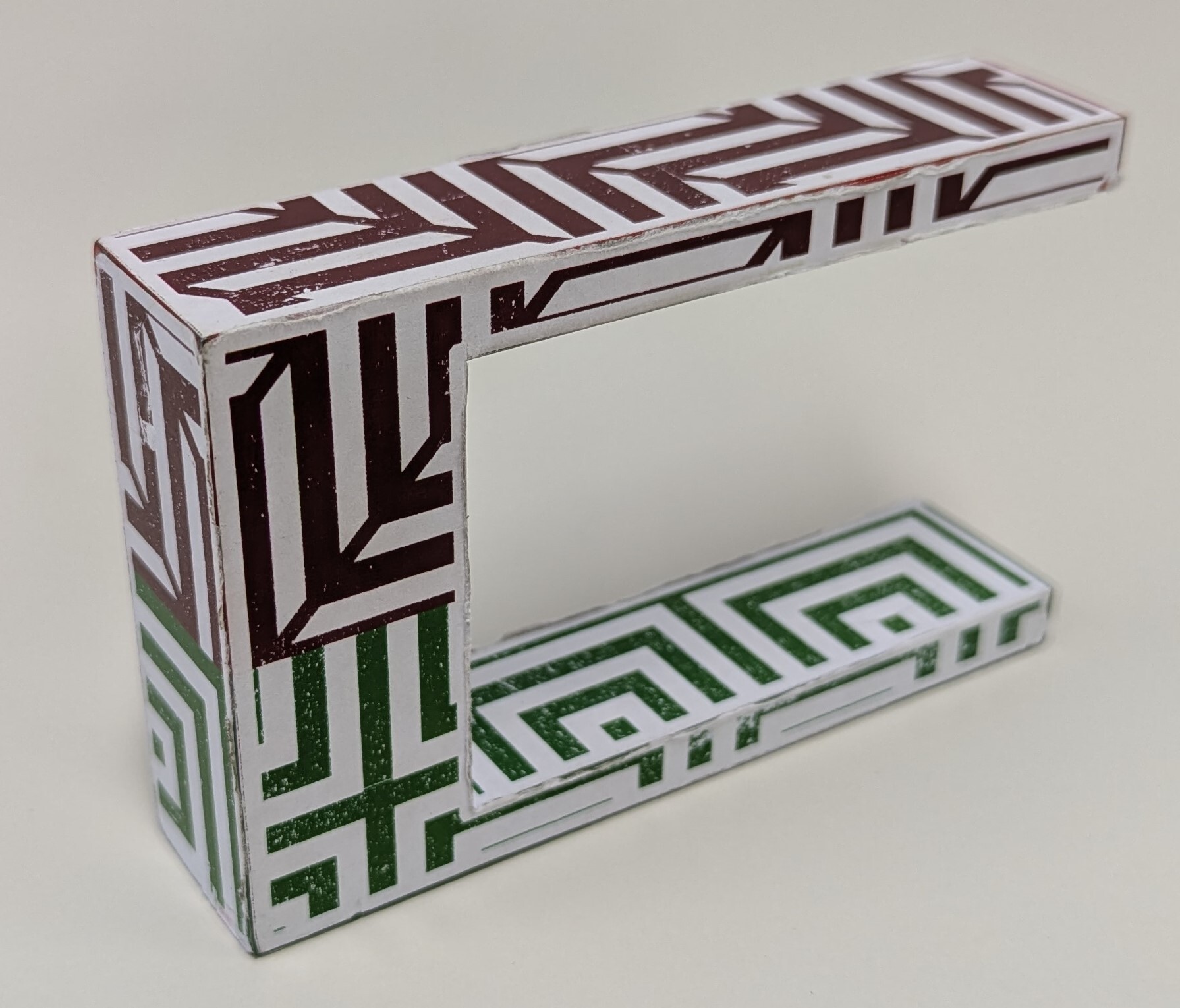}
   \caption
   {The horseshoe magnet with a particular pattern attached to its surface---for better performance of the object tracking.}
   \label{fig:magnet_with_pattern}
\end{figure}

Since the appearance of the representations also depends on the external magnetic field, we use the object tracking capability of the HL2 and let it track the position and orientation of the horseshoe magnet. Because contemporary object tracking technology works reliably mostly for objects with a rich feature pattern, it is technically quite demanding to track the very simple geometric shape of the horseshoe magnet. In order to improve tracking, we have added a non-symmetric color pattern to the physical magnet (and included the same pattern in the 3D model of the magnet that is used in the AR app), as can be seen in \fref{fig:magnet_with_pattern}.

Finally, the magnetic field due to the current through the conductor should be displayed with its symmetry axis aligned with the conductor. Since the conductor is too thin to be tracked directly, we developed a two-step method. First, the position of the fixed conductor mounting is determined at the beginning of the experiment by using image tracking of a sticker placed at one corner of the conductor mounting (see the sticker in \fref{fig:conductorMounting}). Second, the displacement of the conductor from the vertical rest position is measured by a rotary encoder located in the conductor mounting.

\subsubsection{User Interaction with the AR App}
The interaction with the HL2 app is done by using a so-called hand menu, i.e.\ the HL2 recognizes certain hand gestures of the user and thus allows them to be able to press virtual buttons. There is one general menu---the \textit{supervisor menu}---that allows for all possible interactions with the app and another, much more restricted, menu---the \textit{student menu}---that is present for the learners and allows them to interact with the representations. We will briefly describe both menus in the present section.

\begin{figure}[h]
 \centering
  \includegraphics[width=10cm]
      {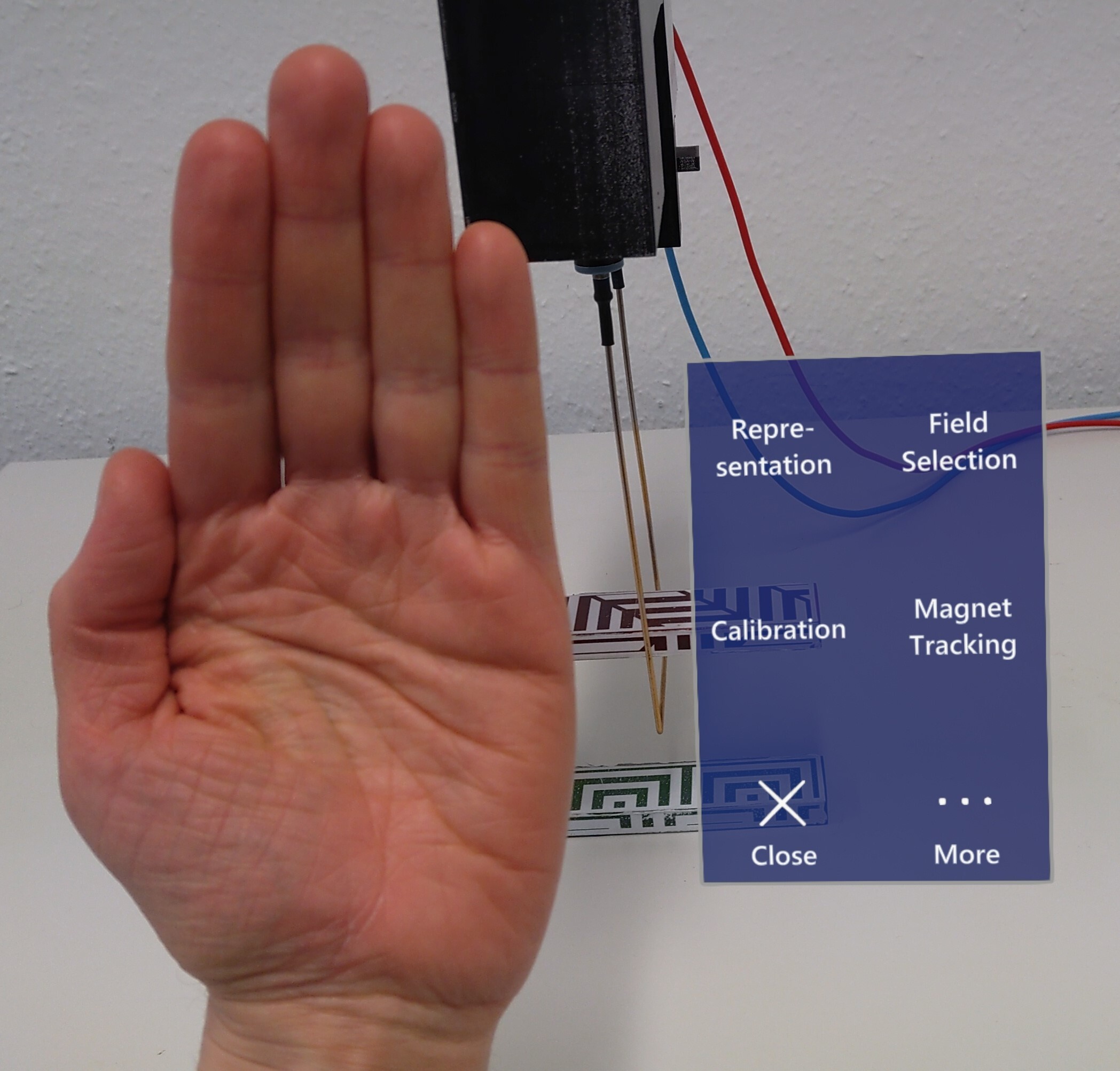}
   \caption
   {The supervisor main menu.}
   \label{fig:sup_main}
\end{figure}

The supervisor menu enables the laboratory supervisor to prepare the app for the student and also to address any technical problems during the experimental phase. It is available from the initial start of the app and is displayed when the user holds up their hand (see \fref{fig:sup_main}). 
The menu contains the options listed in \tref{tab:sup_menu_main}.
\renewcommand{\arraystretch}{1.2}
\begin{table}[h]
\caption{\label{tab:sup_menu_main}Possible options that can be chosen in the supervisor menu.}
\begin{indented}
\item[]\begin{tabular}{@{}lp{0.6\textwidth}}
\br
Menu option&Description\\
\mr
Representation&
\begin{minipage}[t]{0.6\textwidth}
\setstretch{0.5}
Selection of one of the three possible representations: field vector, field line, right-hand rule
\end{minipage} \\
Field selection&
\begin{minipage}[t]{0.6\textwidth}
\setstretch{0.5}
Selection of the magnetic field to be displayed: external, conductor, superposition
\end{minipage}\\
Calibration&
\begin{minipage}[t]{0.6\textwidth}
\setstretch{0.5}
Sub-menu for calibrating the AR setup
\end{minipage}\\
Magnet tracking&
\begin{minipage}[t]{0.6\textwidth}
\setstretch{0.5}
Activate or deactivate the real-time tracking of the horseshoe magnet
\end{minipage}\\
Measurement data&
\begin{minipage}[t]{0.6\textwidth}
\setstretch{0.5}
Display the data that is currently measured
\end{minipage}\\
Settings&
\begin{minipage}[t]{0.6\textwidth}
\setstretch{0.5}
Sub-menu for changing additional settings
\end{minipage}\\
Diagnostics&
\begin{minipage}[t]{0.6\textwidth}
\setstretch{0.5}
Diagnostic tools to check the status of the AR system
\end{minipage}\\
\br
\end{tabular}
\end{indented}
\end{table}

The options \emph{calibration}, \emph{settings} and \emph{diagnostics} lead to further sub-menus. In the calibration sub-menu we can choose to set the position of the conductor mounting and calibrate the current measurement. We also added the possibility to deactivate the real-time magnet tracking and choose predefined magnet positions. In this way, we are more flexible in cases where the tracking quality is diminished, e.g.\ in situations with inadequate light sources.

In the settings sub-menu, we can change additional, advanced settings, such as choosing which conductor mounting to connect to (in situations where several AR setups are operating next to each other). 

Finally, there is a diagnostics sub-menu that allows the supervisor to monitor the correct functioning of the app, such as checking if the Bluetooth connection is active.

After the setup phase is done, the supervisor can close the supervisor menu and hand over the AR smartglasses to the learners. The learners can now use the student menu that enables them to switch between the representations, the displayed magnetic fields and, in case the real-time tracking of the horseshoe magnet has been deactivated, choose between the predefined magnet positions.

Additionally, the AR app includes the possibility to wirelessly communicate with external devices, like laptops or tablets, through a web-interface (based on the work presented in \cite{kapp2021arett}). The latter can be used by an instructor to manipulate---from outside the AR app---the visualizations shown in the app.

\subsection{3D Representations in AR} \label{sec:reps_3d}
\begin{figure}[h!]
    \centering
    \begin{adjustbox}{max width=\textwidth}
        \begin{minipage}{0.49\textwidth}
            \centering
            \subcaption{}
            \includegraphics[width=\linewidth]{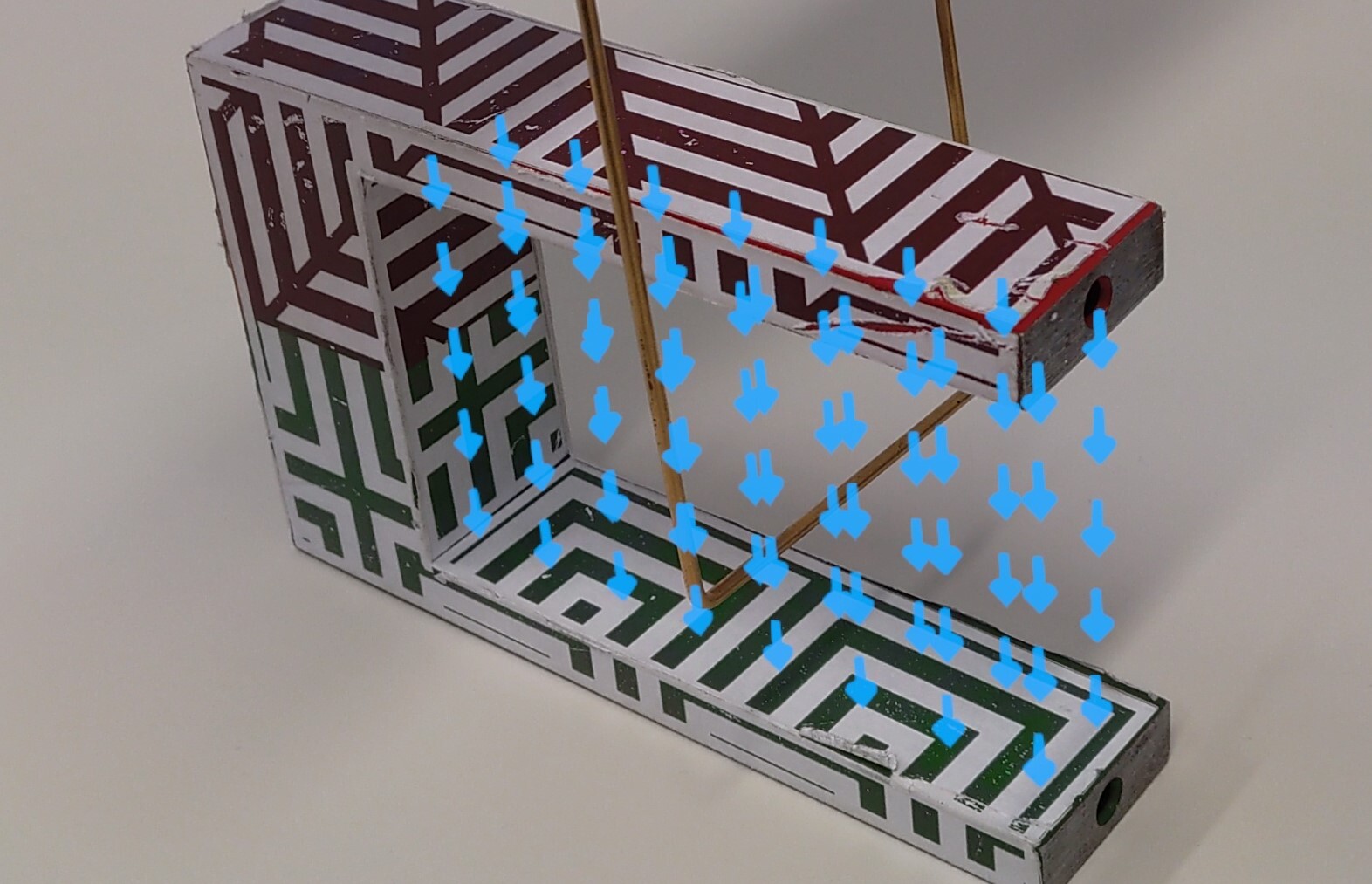}   
        \end{minipage}
        \hspace{-0.5em}
        \begin{minipage}{0.49\textwidth}
            \centering
             \subcaption{}
            \includegraphics[width=\linewidth]{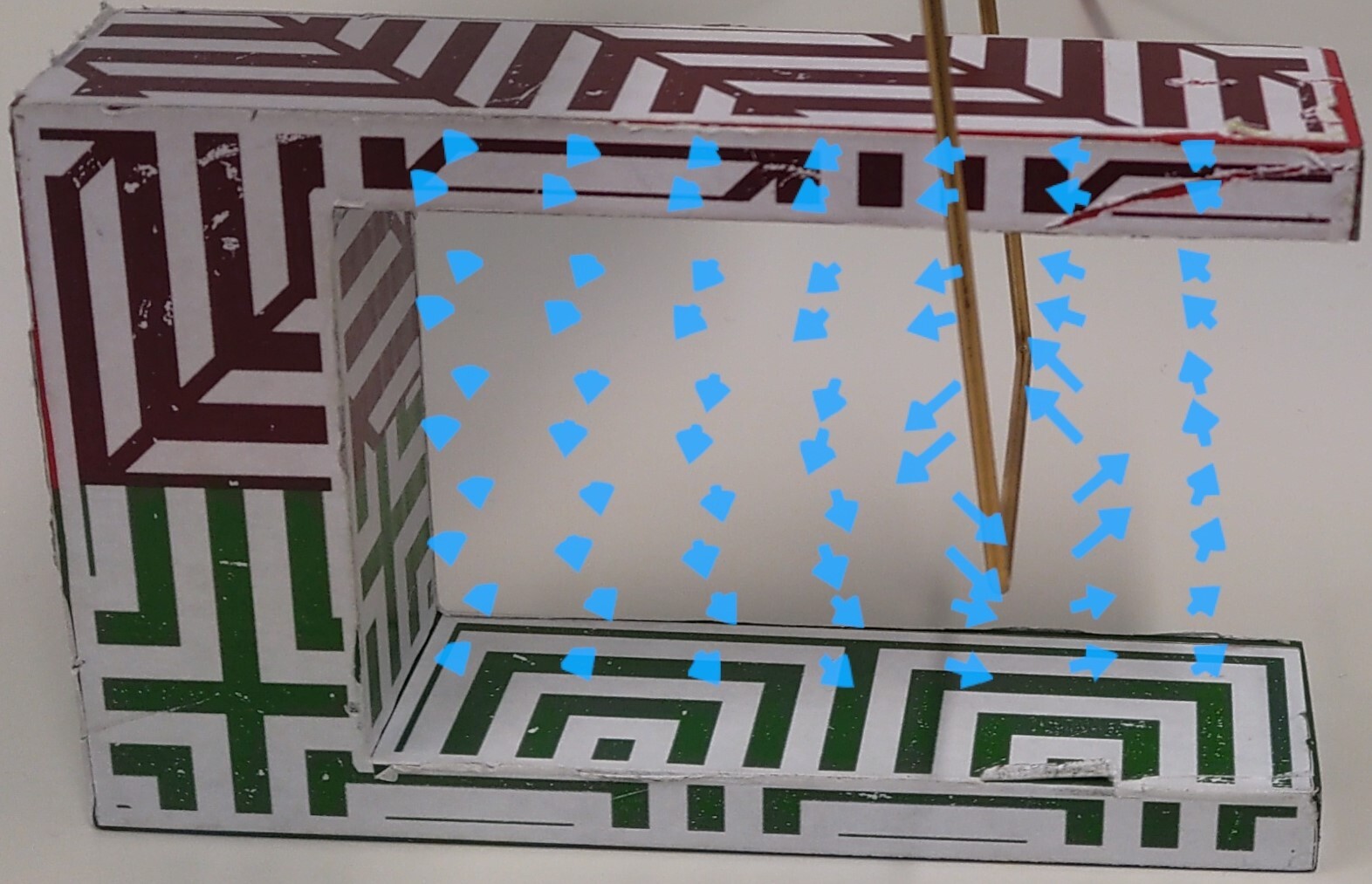}       
        \end{minipage}
    \end{adjustbox}
    \begin{minipage}{0.49\textwidth}
        \centering
        \includegraphics[width=\linewidth]{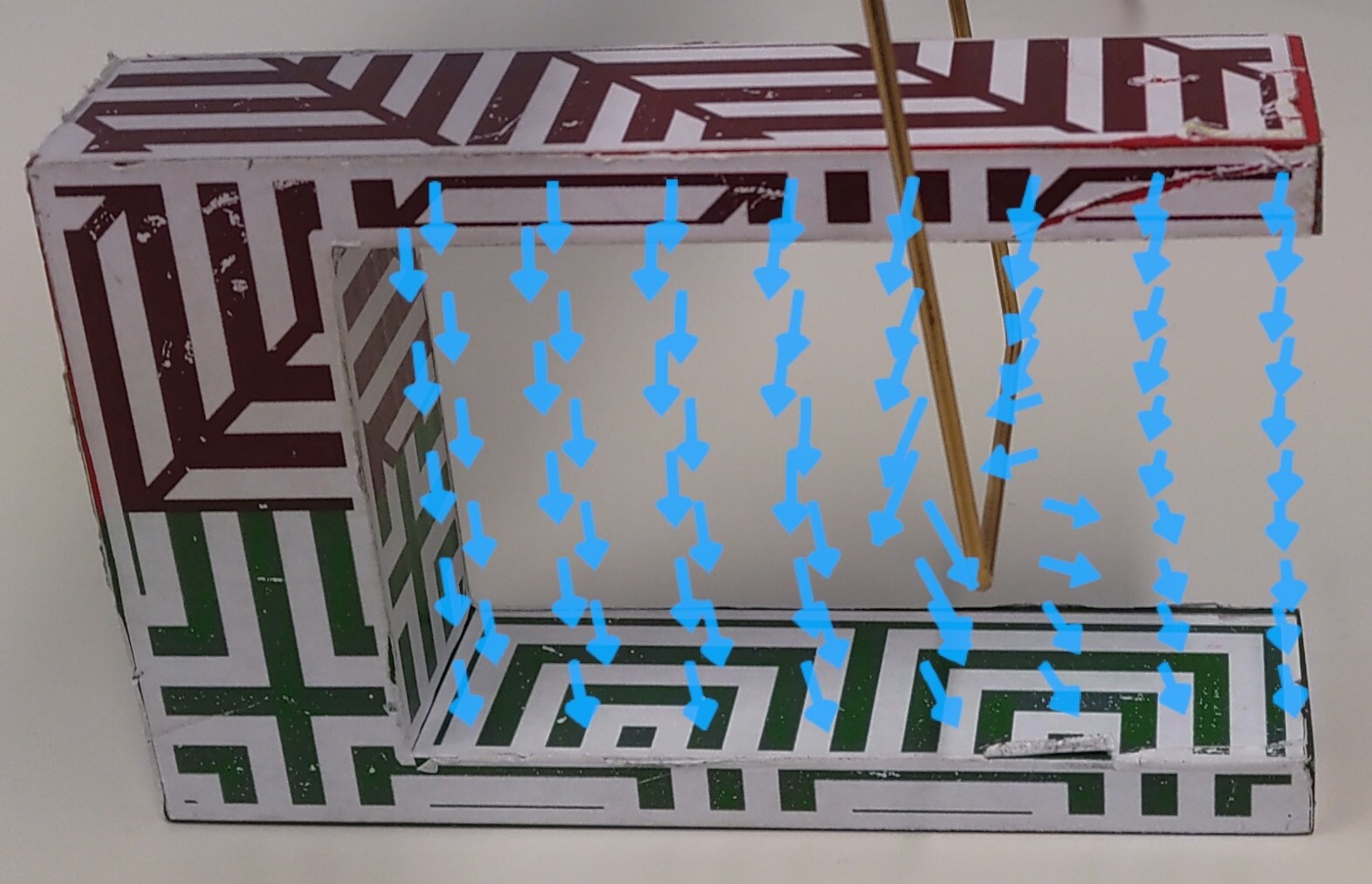}
        \subcaption{}
    \end{minipage}
    \caption{The 3D field vector representations of a) the external magnetic field, b) the conductor's magnetic field and c) the total magnetic field. Although in these images the arrows occlude the real-world objects, in reality, they are semi-transparent.}
    \label{fig:rep_3D_vector}
\end{figure}

In the following, we present the 3D representations that can be displayed in the AR environment. In \fref{fig:rep_3D_vector} we show the field vector representations for the three different magnetic fields.

\begin{figure}[h!]
    \centering
    \begin{adjustbox}{max width=\textwidth}
        \begin{minipage}{0.475\textwidth}
            \centering
            \subcaption{}
            \includegraphics[width=\linewidth]{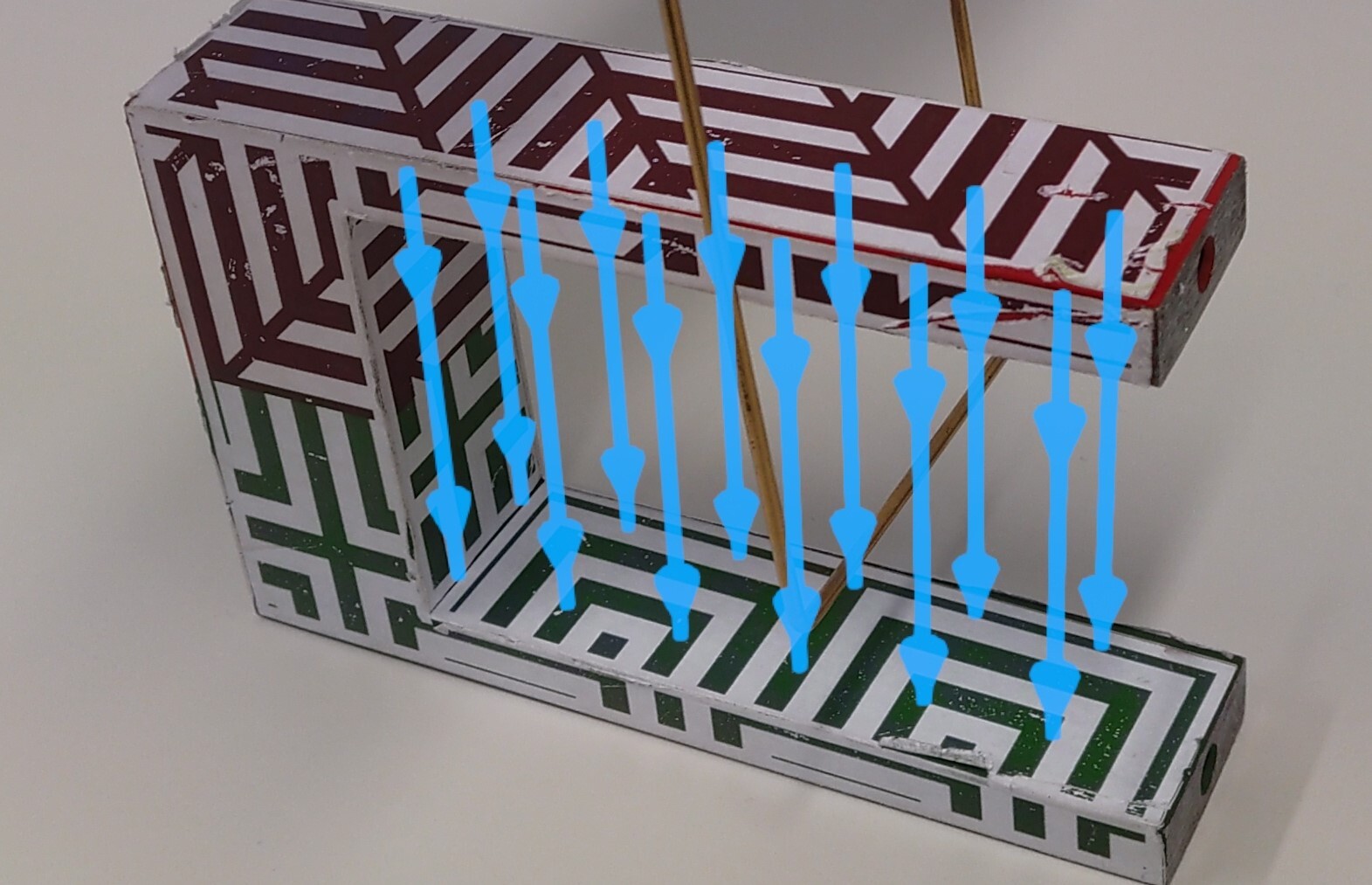}   
        \end{minipage}
        \hspace{-0.5em}
        \begin{minipage}{0.475\textwidth}
            \centering
             \subcaption{}
            \includegraphics[width=\linewidth]{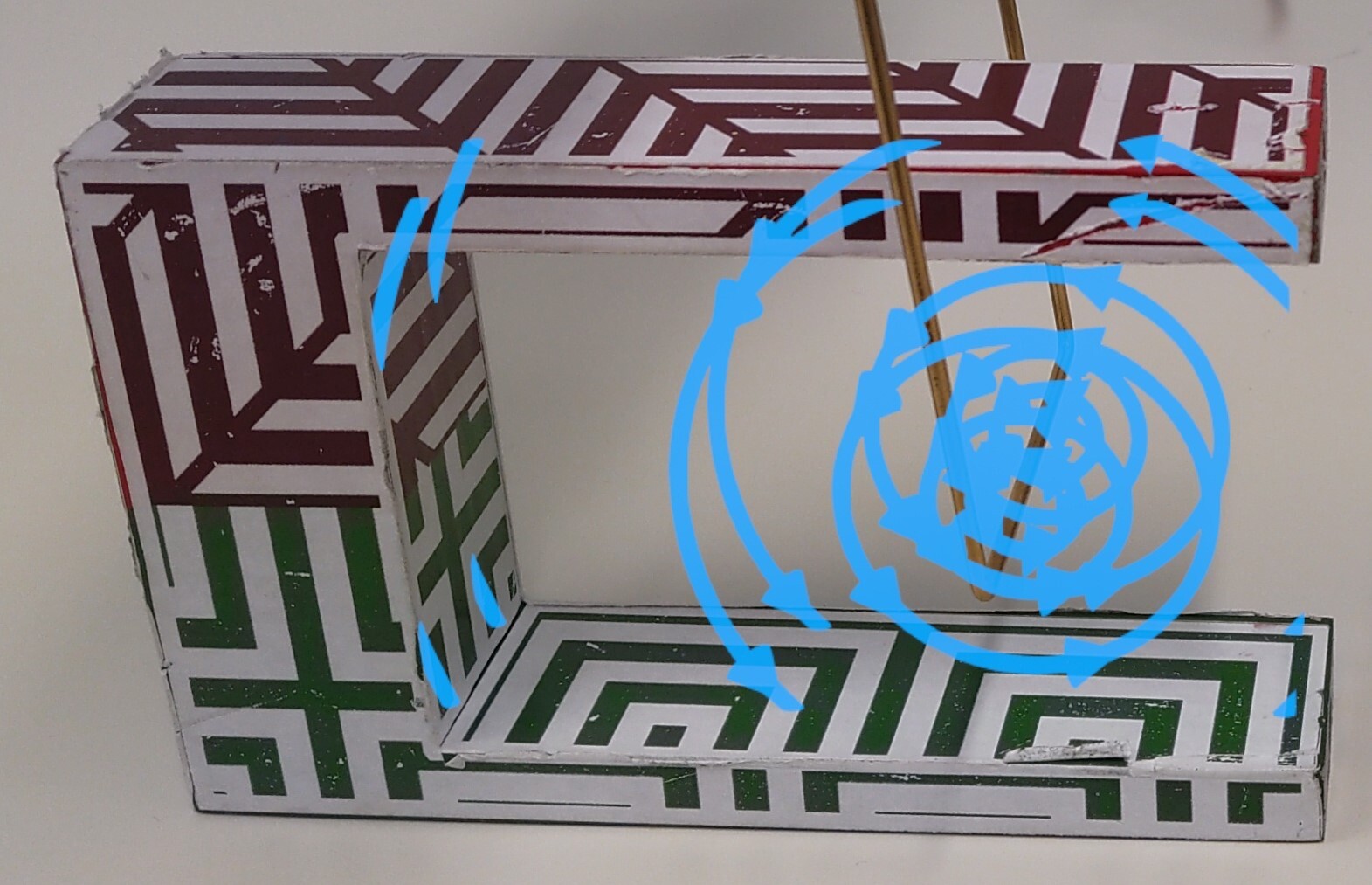}       
        \end{minipage}
    \end{adjustbox}
    \begin{minipage}{0.475\textwidth}
        \centering
        \includegraphics[width=\linewidth]{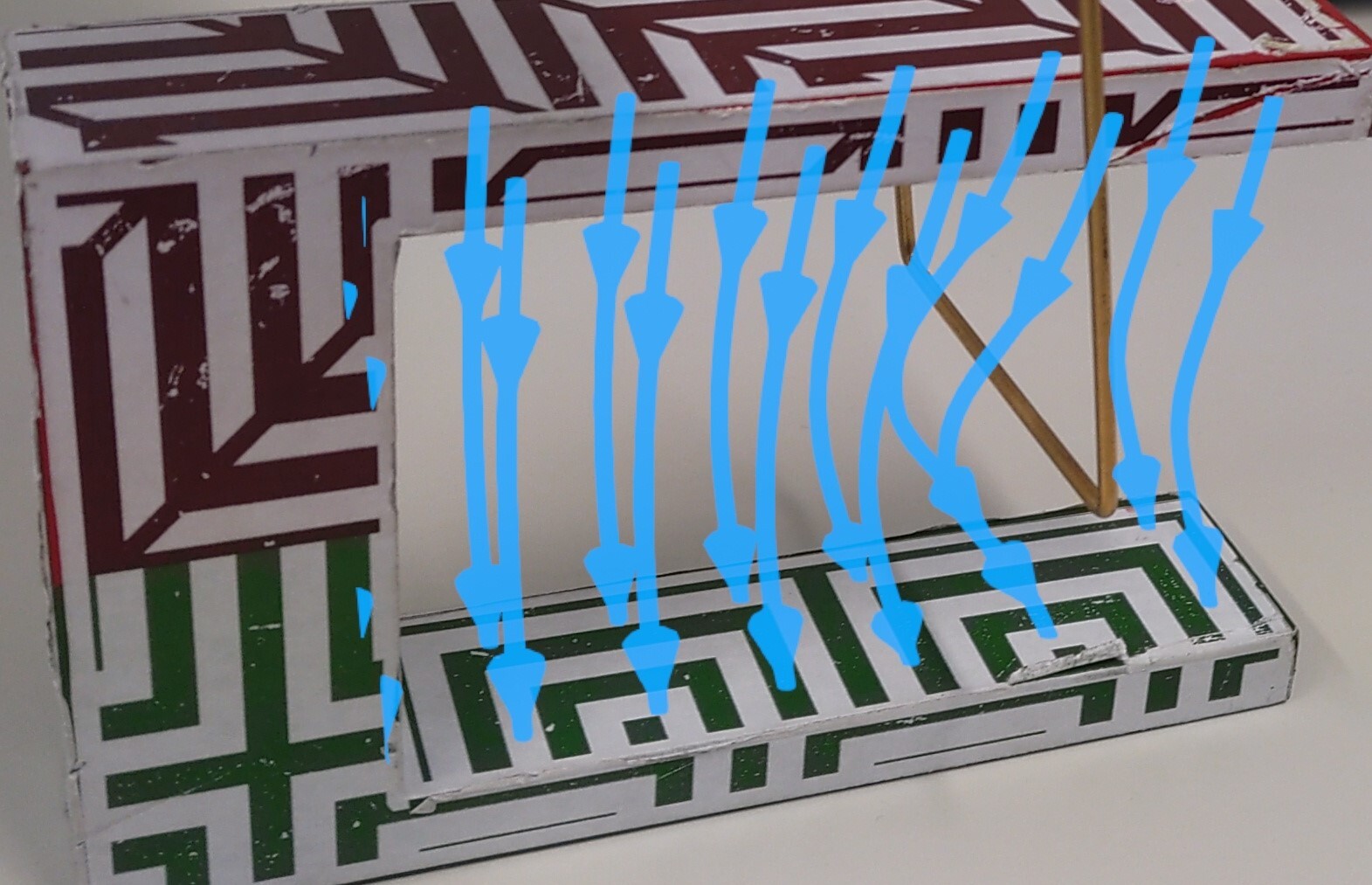}
        \subcaption{}
    \end{minipage}
    \caption{The 3D field line representations of a) the external magnetic field, b) the conductor's magnetic field and c) the total magnetic field. Although in these images the field lines occlude the real-world objects, in reality, they are semi-transparent.}
    \label{fig:rep_3D_line}
\end{figure}

The field line representations are displayed in \fref{fig:rep_3D_line}.
Note that for educational purposes, the visualizations of the conductor's magnetic field and its superposition with the external magnetic field only take into account the idealized conductor's magnetic field \eref{eq:mag_field_cond}, i.e.\ only the field due to the horizontal segment of the conductor swing (see also \sref{sec:exact_conductor_field}). Thus, we keep the focus on learning about the magnetic field of a straight wire.

\begin{figure}[h!]
 \centering
  \includegraphics[width=0.6\textwidth]
      {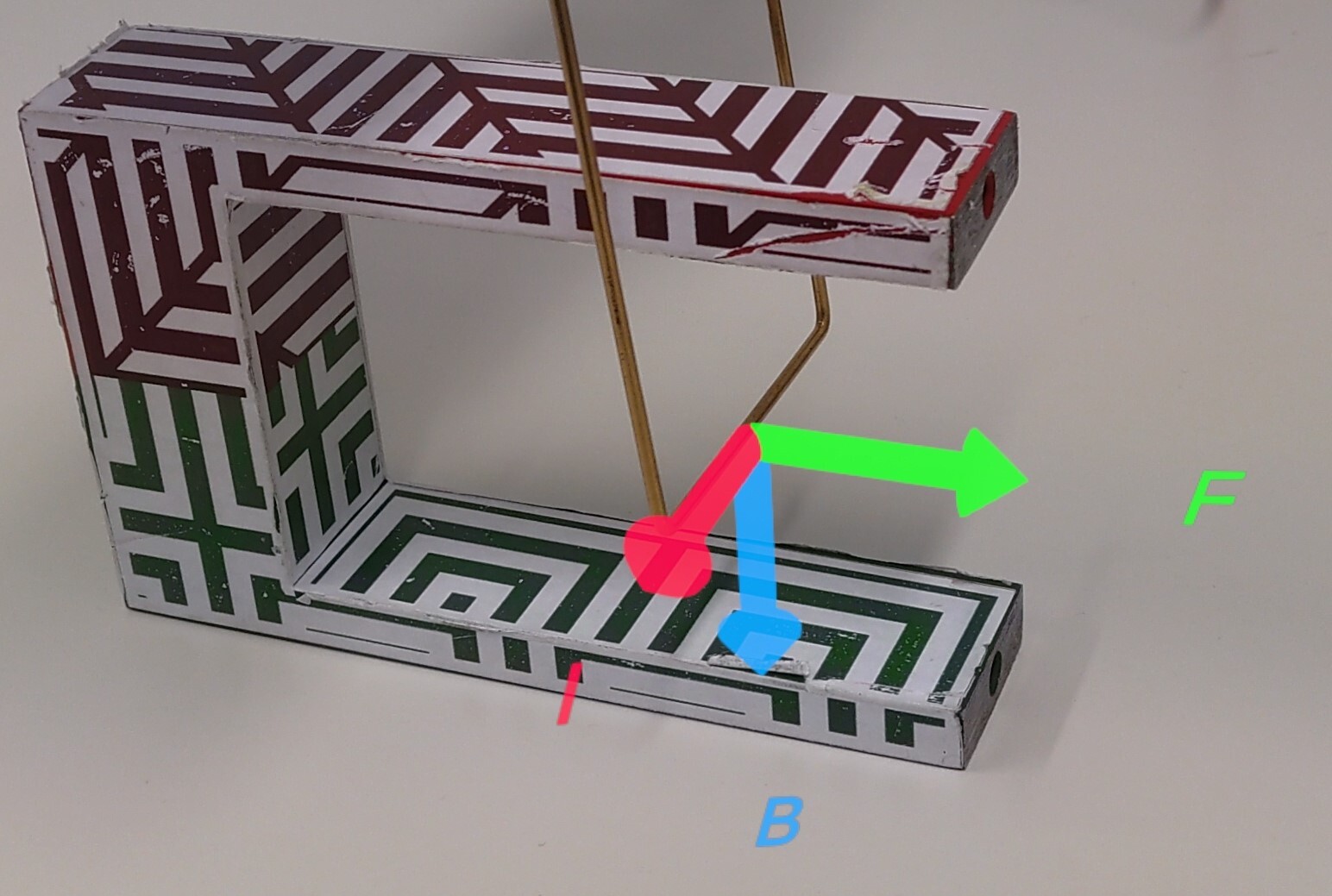}
   \caption
   {The 3D right-hand rule representation with arrow labels in the form of mathematical symbols for the physical quantities.}
   \label{fig:rep_3D_tripod}
\end{figure}

The right-hand rule representation is shown in \fref{fig:rep_3D_tripod}.

\subsection{Evaluation of the System Usability}
We evaluated the usability of the AR learning environment by handing out questionnaires to 188 high school students, using the system usability scale from \cite{brooke1996sus}. The scale led to an average score of 80.18 (SD: 12.06) which can be associated with an assessment of ``excellent'' according to \cite{bangor2009determining}. This finding suggests that the augmented experimental setup is easy to use, enabling learners to gain a deeper understanding of the Lorentz force and representational competence.

\section{Conclusion and Outlook} \label{sec:conclusion}
We presented a novel AR environment for learning the concept of the Lorentz force and the superposition of magnetic fields. The AR setup consists of the conventional conductor swing experiment that is enhanced by the usage of AR smartglasses. These enable the display of MERs that react dynamically in real time on the learners' manipulations. The AR setup involves the precise measurement of the conductor current and the resulting displacement of the conductor swing.

We are currently systematically investigating the learning effectiveness of this AR learning environment. Results from those studies will be reported in due course. 

In the future, the AR app could be easily extended by the inclusion of additional representations (e.g.\ a streamline or needles representation), the visualization of more complex magnetic field structures (e.g.\ the more realistic magnetic field of the horseshoe magnet) and additional measurement sensors, such as a force sensor to measure the magnitude of the Lorentz force directly.

We believe that the present AR setup is well-suited to provide a powerful learning framework for both basic concepts in classroom settings---like the display of the right-hand rule--- and more advanced phenomena---like visualizing complex magnetic fields---for physics students in laboratory courses. Using an external device, such as a laptop, it is possible for an instructor to directly manipulate the AR app that is running on the students' AR smartglasses. Thus, the presented AR learning environment is a valuable asset for teachers that provide inquiry-based learning in electromagnetism.

\ack
We would like to express our gratitude to the \emph{CESAR} project team for their valuable support. In particular, we thank Zoya Kozlova, Roman Schmid, Roland Brünken, Peter Edelsbrunner, Sarah Hofer, Sarah Malone, Ralph Schumacher, Elsbeth Stern and Andreas Vaterlaus for their contributions to data collection, analysis, and conceptual discussions and we greatly appreciate their collaboration.

This work was funded by the Deutsche Forschungsgemeinschaft (DFG, German Research Foundation)---Projektnummer 471917560---and the Schweizerischer Nationalfonds (SNF, Swiss National Science Foundation)---Projektnummer 10019L\_204987.

We would also like to thank Marcus Hauser from \emph{Bomatec AG} for providing useful insight into the composition of horseshoe magnets, and Matthias Mitschele for creating and thoroughly testing the magnet surface pattern to improve object track-ability.

\section*{References}
\bibliographystyle{unsrt.bst} 
\bibliography{bibliography}  

\end{document}